\documentclass[pre,twocolumn,showpacs,preprintnumbers,superscriptaddress,amsmath,amssymb]{revtex4}
\usepackage{graphicx}
\usepackage{amssymb}
\usepackage{amsmath}
\usepackage{amsfonts}
\usepackage{mathrsfs}
\usepackage{subfigure}
\usepackage{color}
\usepackage[colorlinks,linkcolor=blue,citecolor=blue]{hyperref}
\usepackage{subfigure}

\begin{document}
\title{Fluctuation Theorems Containing Information for Autonomous Maxwell's Demon-assisted Machines}
\author{Xuehao Ding}
\affiliation{School of Physics, Peking University, Beijing 100871, China}
\author{H. T. Quan}\email[]{htquan@pku.edu.cn}
\affiliation{School of Physics, Peking University, Beijing 100871, China}
\affiliation{Collaborative Innovation Center of Quantum Matter, Beijing 100871, China}

\begin{abstract}
In this article, we introduce two kinds of Fluctuation Theorems (FT) containing information for autonomous Maxwell's demon-assisted machines. Using Jensen's Inequality, we obtain Landauer's principle formulation of the second law for the whole process of the machine. Finally we make use of our results to analyze a new information device.
\end{abstract}
\maketitle

\section{INTRODUCTION}
In 1871~\cite{Maxwell1871}, James C. Maxwell conceiced an intelligent creature, now known as Maxwell's demon~\cite{Leff1992Maxwell} to challenge the second law of thermodynamics. In 1982, based on Landauer's princeple~\cite{Landauer1961IBM}, C. H. Bennett gave an explanation~\cite{Bennett1982IJTP} of Maxwell's demon, involving the concept of the information entropy (or Shannon entropy), and finally solved the lone-time dispute. The study of Maxwell's demon has triggered a lot of interests in the exploration of physics of information \cite{Quan2006PRL,Jordan2010PRE,Vaikuntanathan2011PRE,Abreu2012PRL,Deffner2013PRX,Barato2014PRL,Lu2014physicstoday,Jordan2014PRX,Parrondo2015Nature,Pekola2015Nature,Kutvonen2016PRE,Shubhashis2016,Merhav2017,Boyd2017PRE,Strasberg2017PRX}.

\begin{figure}[!htb]	
	\includegraphics[width=0.5\textwidth,clip]{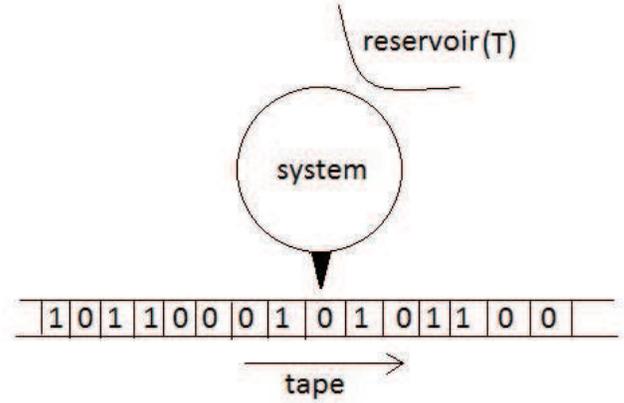}
	
	\caption{\label{fig1}Schematic plot of the Maxwell's demon-assisted machine. The tape represents the memory in Maxwell's demon.}
\end{figure}

Following the understanding of C. H. Bennett, we can simplify the autonomous Maxwell's demon-assisted engine~\cite{Mandal2012PNAS} to a system coupled to a heat bath and a tape (see Fig. \ref{fig1}). The tape acts as the Maxwell's demon's memory. When the machine converts heat into work, information will be written to the tape, or conversely when work is cost and converted into heat, the information on the tape will be (partially) erased. In the former case, the sum of the entropy production of the system and the heat bath can be negative, which is compensated by the increase of the information entropy of the tape. Along this line, some exactly solvable models of Maxwell's demon-assisted machine, such as the information refrigerator~\cite{Mandal2013PRL}, the information pump~\cite{Cao2015PRE} are proposed. Under some conditions, they can also work as information erasers. The evolution of the composite system (the system plus the bit flow) is governed by a Master equation, and is periodically interrupted by new coming bits. It has been demonstrated that the second law of thermodynamics is valid in these autonomous information machines~\cite{Mandal2012PNAS, Mandal2013PRL, Cao2015PRE} as long as the information entropy of the bits is put on the same footing as thermodynamic entropy. However, a detailed understanding of these fluctuating quantities, such as information, heat and work in these machines is still lacking. It is thus desirable to explore stronger relations among these quantities, such as Fluctuation Theorems in these information machines.

In 2005, Udo Seifert introduced the entropy for the microstate~\cite{Seifert2005PRL} in stochastic thermodynamics, that is $s(\Gamma)=-\ln[p(\Gamma)]$, where $\Gamma$ denotes the microstate of the system, and $p(\Gamma)$ denotes the probability distribution of the microstate $\Gamma$. Using this definition, Seifert derived an Integral Fluctuation Theorem (IFT) for stochastic processes,
\begin{equation}\label{equ2}
\left<e^{-\Delta s_{tot}}\right>=1,
\end{equation}
where $\Delta s_{tot}=\Delta s + \Delta s_{bath}$ is the fluctuating total entropy change, and $\Delta s_{bath}$ is the fluctuating entropy change of the heat bath. Using Jensen's Inequality, one can immediately obtain the second law of thermodynamics,
\begin{equation}\label{equ3}
	\Delta S_{tot}\equiv \left<\Delta s_{tot}\right>\geq 0,
\end{equation}
where $\Delta S_{tot}=\Delta S+\Delta S_{bath}$ is the total entropy change, $\Delta S=<\Delta s>$ and $\Delta S_{bath}=<\Delta s_{bath}>$ are the ensemble averages of the fluctuating entropy changes.

However, the IFT Eq.(\ref{equ2}) can not be directly applied to the information machine. Because the IFT Eq.(\ref{equ2}) is valid when the evolution of the system is governed by a Master equation for a single period (see, e.g., Refs.~\cite{Jarzynski1999PRE,Min2005Nano,Seifert2005EPL,Lacoste2008PRE}), but the evolution of the information machine consists of many periods which are interrupted by new coming bits. The evolution of the composite system cannot be reduced to a process governed by a single Master equation. For this reason, FTs describing the information machine are still lacking. In this article, we derive the FTs containing information for a class of information machines. Using the IFTs, we can derive the second law for the whole process. Our FTs containing information content are different from the FT with measurement and feedback \cite{Sagawa2010PRL,Sagawa2012PRL} in three aspects. First, we use the information entropy to characterize the information content, while in the FT with measurement and feedback, the information content is characterized by the mutual entropy between the system and the demon's memory. Second, similar to Seifert's FT Eq.(\ref{equ2}), the FT with measurement and feedback \cite{Sagawa2010PRL,Sagawa2012PRL} is also valid for a single process governed by a Master equation but does not apply to information machines because the evolution consists of many periods. Third, the information machine is autonomous. There is no measurements and feedback controls. Furthermore, we make use of our results to analyze a new information device~\cite{Mcgrath2016PRL}.

This paper is organized as follows: In Sec. II we derive the IFTs, the Detailed Fluctuation Theorems (DFT) and the second law for this class of models. In Sec. III We use our methods to study the IFT for a new model proposed recently by Thomas McGrath, et al. In Sec. IV we give some discussions and summarize our paper.
\section{FLUCTUATION THEOREMS FOR INFORMATION MACHINES}
Let us recall the models of information machine proposed in Refs.~\cite{Mandal2012PNAS,Mandal2013PRL,Cao2015PRE}. In these three models, the information engine, the information pump and the information refrigerator, there is a system and a tape. Every bit of the tape interacts with the system for a fixed time interval t consecutively. There is only one bit interacting with the system at a time.

\begin{figure}[!htb]
	\subfigure[information engine]
	{
		\includegraphics[width=0.1725\textwidth]{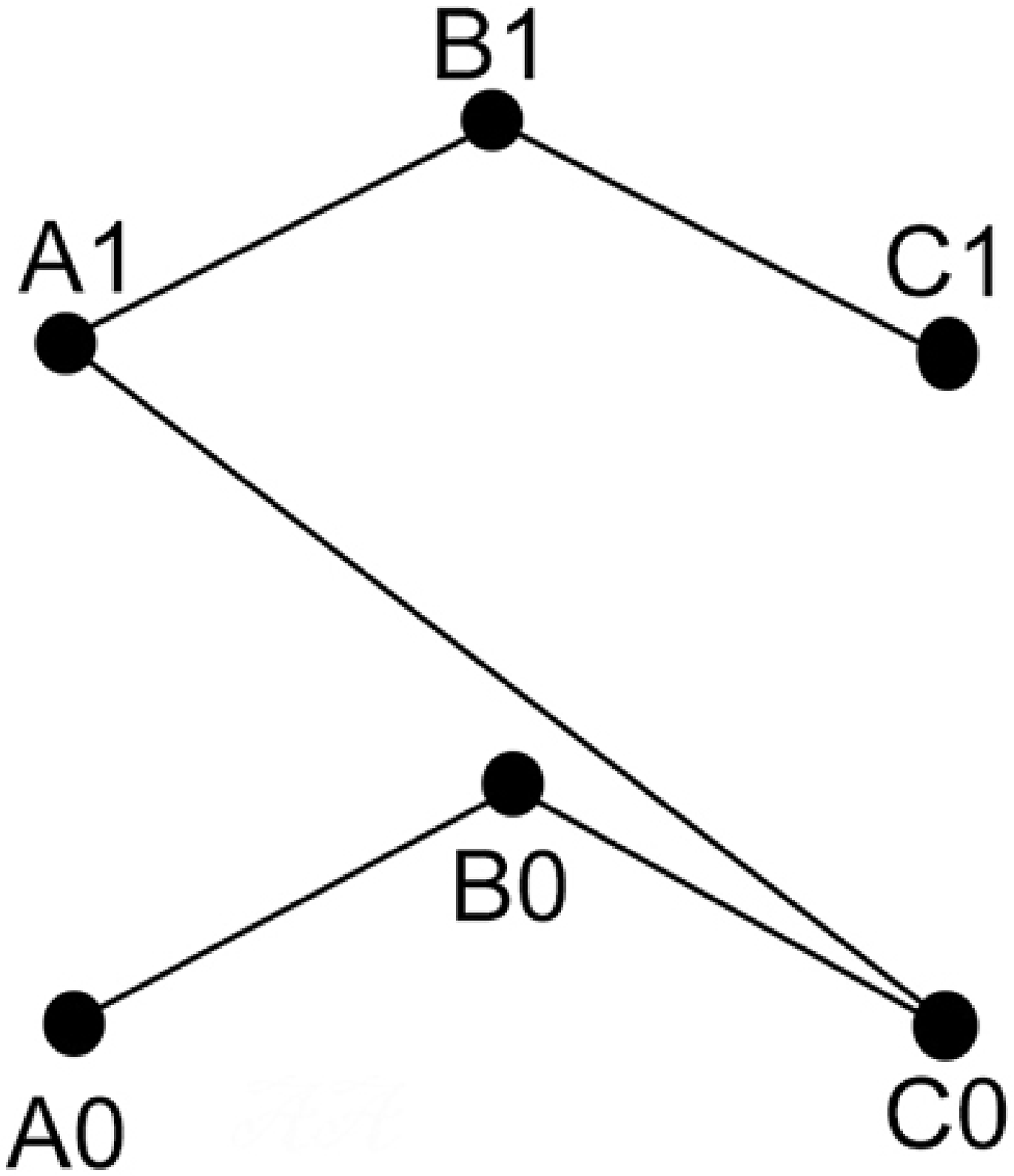}\label{fig2:a}
	}
    \subfigure[information pump]
    {
		\includegraphics[width=0.2\textwidth]{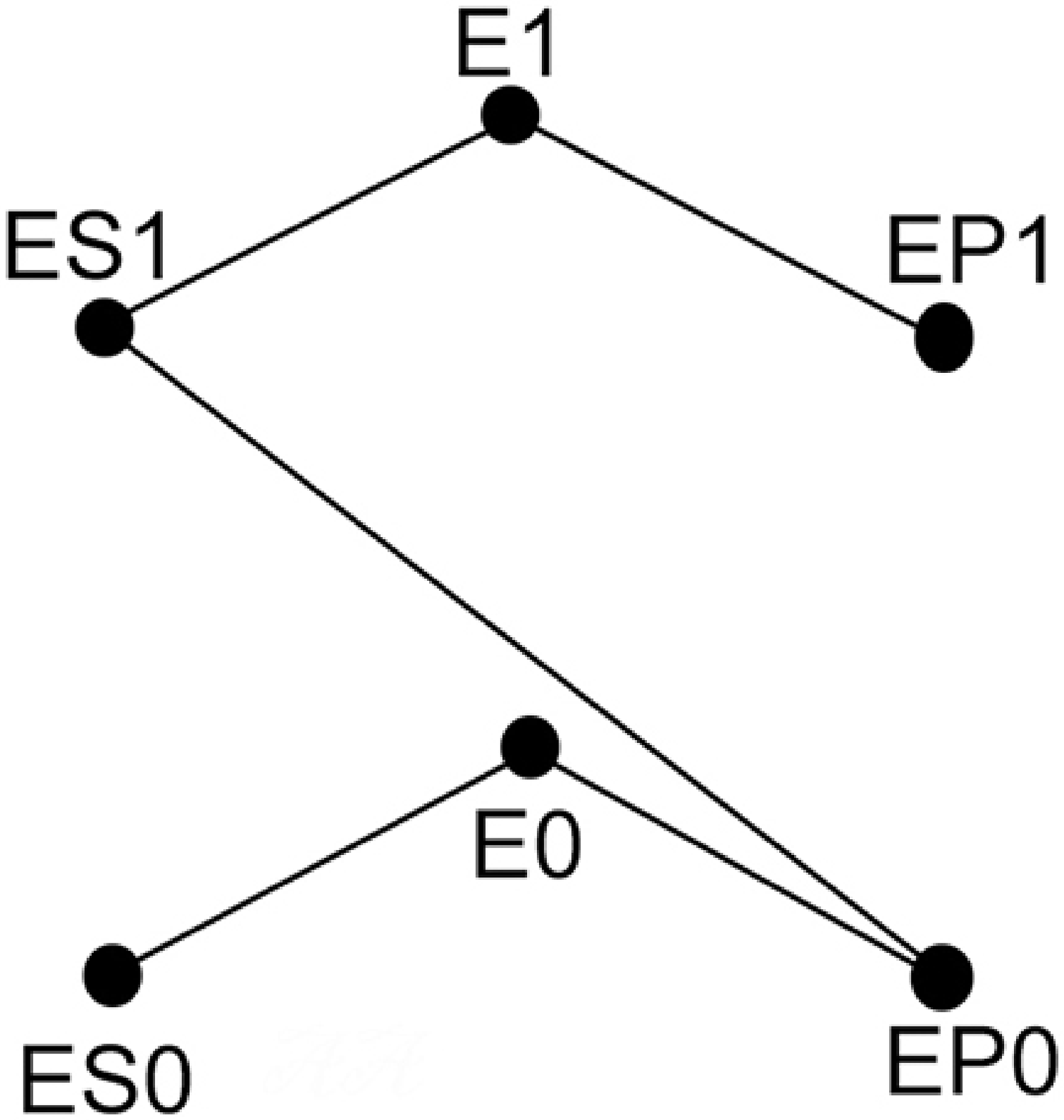}\label{fig2:b}
	}
	\caption{Schematic plots of the information engine and the information pump. (a) is the information engine~\cite{Mandal2012PNAS}, (b) is the information pump~\cite{Cao2015PRE}. A, B and C (E, EP and ES) represent the 3 states of the system. 0 and 1 denote the two states of the bit. For the information engine, the entropy production of the environment can be expressed as $\Delta S_{bath}=-\frac{W}{T}$, where W is the work the engine does. For the information pump, the entropy production of the environment can be expressed as $\Delta S_{bath}=\frac{\Delta n\cdot\Delta \mu}{T}$, where $\Delta n$ is the number of moleculars converted form EP to ES, and $\Delta \mu$ is the chemical potential difference between EP and ES.}
	\label{fig2}
\end{figure}

The information engine and the information pump are very similar. The system itself has three states. And when the system is coupled to a bit, the whole system can be viewed as a composite system of six states (see Fig. \ref{fig2}). The allowed transitions between these states are illustrated with solid lines in Fig. \ref{fig2}. When a transition between A (ES) and C (EP) occurs, an accompanied flip of the bit will occur. Meanwhile, for the information engine energy will flow into (out of) the work reservoir, and for the information pump a chemical reaction will occur.

\begin{figure}[!htb]
	\centering	
	\includegraphics[width=0.2\textwidth]{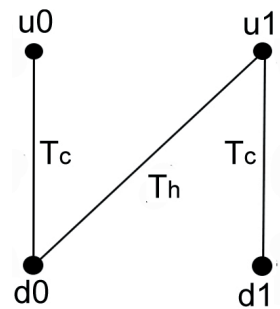}
	\caption{Schematic plot of the information refrigerator~\cite{Mandal2013PRL}. u and d denote the two states of the system with different energies. 0 and 1 denote the two states of the bit. $T_h$ and $T_c$ denote the temperatures of the hot heat bath and the cold heat bath. For the information refrigerator, the entropy production of the environment can be expressed as $\Delta S_{bath}=\frac{Q_c}{T_c}+\frac{Q_h}{T_h}$, where $Q_c$ and $Q_h$ denote the heat transported into the cold heat bath and the hot heat bath respectively.}
	\label{fig3}
\end{figure}

The information refrigerator is slightly different from the information engine and the information pump. The system itself has two states with different energies. When coupled to a bit, the system can be viewed as a composite system of four states (see Fig. \ref{fig3}). The transition between $d0$ and $u1$ is due to its coupling to a hot heat bath, while the other transitions are due to the coupling to a cold heat bath. When a transition between $d0$ and $u1$ occurs, the bit will flip.

We can summarize this class of models of information machine as follows: First, the system and the incoming bit form a composite system. Second, coupled to one or more heat baths, the whole system evolves under a Master equation for a time interval t. Third, at the end of the period, the evolution ends. Fourth, a new period starts, the initial state of the system is the final state of the system of the last period, but the initial state of the bit probably changes because a new incoming bit replaces the old one. The machine repeats these steps.

During an interacting interval, the probability distribution of the states of the composite system is governed by the Master equation,
\begin{equation}\label{equ4}
	\dot{p_{i}}=\sum_{j}p_{j}W_{ij},
\end{equation}
where $W_{ij}$ is the transition rate matrix between the states of the composite system~\cite{Mandal2012PNAS, Mandal2013PRL, Cao2015PRE}.

For the convenience of the following analysis, we introduce the following notations. $S_x$ denotes the entropy of the system, $I_b$ denotes the information entropy of the bit, $M$ denotes the mutual entropy between the system and the bit, $S_{bath}$ denotes the entropy of the heat bath, $p(0)$ ($p(1)$) denotes the probability of state 0 (1) of the incoming bit.
\subsection{DERIVATION AND DISCUSSIONS OF THE IFTS}

For the definition of the information entropy, we adapt base-e logarithm instead of base-2 logarithm. Then the information entropy $I_b$ has the same form as the thermodynamic entropy (we set Boltzmann constant $k_B$=1 in this article),
\begin{equation}\label{equ5}
	I_b\equiv -\sum_{b=0,1}p(b)\cdot \ln [p(b)].
\end{equation}
By employing the definition of the entropy for the microstate, we can define the fluctuating information entropy of a bit and the fluctuating mutual entropy between the system and the bit,
\begin{equation}\label{equ6}
	\mathscr{I}(b)\equiv-\ln[p(b)],
\end{equation}
\begin{equation}\label{equ7}
	\mathscr{M}\left(x,b\right)\equiv \ln[\frac{p\left(x,b\right)}{p\left(x\right)p\left(b\right)}].
\end{equation}
The ensemble averages of these fluctuating quantities correspond to the information  entropy and the mutual entropy,
\begin{equation}\label{equ8}
	I_b=\left<\mathscr{I}\right>,
\end{equation}
\begin{equation}\label{equ9}
M=\left<\mathscr{M}\right>.
\end{equation}

\begin{figure*}[!htb]
	\centering	
	\includegraphics[width=1\textwidth]{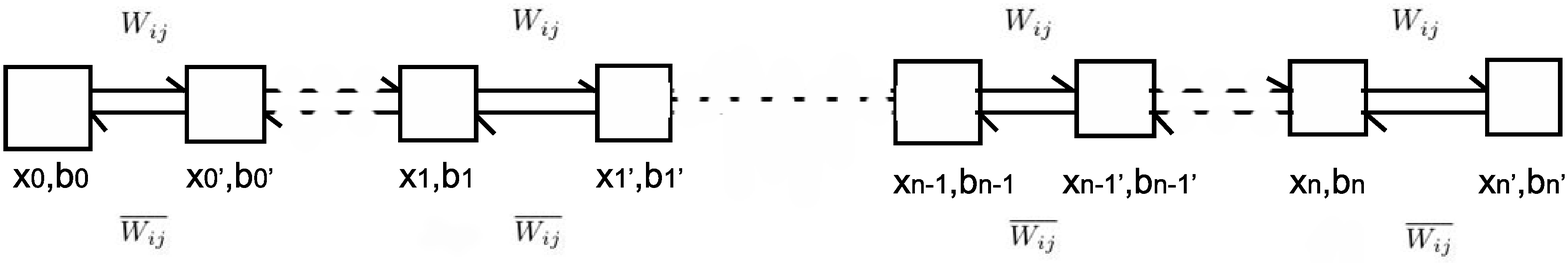}
	\caption{Schematic plot of the whole process of the information machine. Every solid arrow represents the evolution of the whole system during every interacting time interval, and every dashed arrow represents a process of replacing the old bit with the incoming new bit. $(x_i,b_i)$ and $(x_i',b_i')$ denote the initial and the final states of the system and bit in the i-th period. $W_{ij}$ and $\overline{W_{ij}}$ denote the transition rate matrices between the states of the composite system in the forward and the backward processes.}
	\label{fig4}
\end{figure*}

Now we consider a trajectory and its time-reversed trajectory. We set the initial distribution of the backward process as the final distribution of the forward process. In Fig. \ref{fig4}, every solid arrow represents the evolution of the whole system during every interacting time interval, and every dashed arrow represents a process of replacing the old bit with the incoming new bit. For the forward process, $W_{ij}$ is the transition rate matrix between the states (see example Eq.(\ref{equ22})). We use $(x_i,b_i)$ and $(x_i',b_i')$ to denote the initial and the final states of the system and bit in the i-th period. At the end of every period, the state of the system $x_i'$ is recycled as the initial state of the system for the next period $x_{i+1}=x_i'$. The bit is replaced by a new bit, and the initial state of the new bit is randomly sampled from the given distribution of the incoming bit flow $P_F^i(b_i)=p(b_i)$, where $P_F^i(b_i)$ denotes the initial distribution of the bit in the i-th period in the forward process. If we use $(x_0,b_0)$, $(x_0',b_0')$, $(x_1,b_1)$, $(x_1',b_1')$.....$(x_n,b_n)$, $(x_n',b_n')$ to denote a forward trajectory. Its time-reversed trajectory in the backward process is $(x_n',b_n')$, $(x_n,b_n)$, $(x_{n-1}',b_{n-1}')$, $(x_{n-1},b_{n-1})$.....$(x_0',b_0')$, $(x_0,b_0)$. For the backward process, the transition rate matrix is denoted as $\overline{W_{ij}}$, which is identical to $W_{ij}$, i.e., $\overline{W_{ij}}=W_{ij}$ in our paper. Now we can discuss two different choices of the initial distribution of every bit in the backward process. They correspond to $P_B^i(b_i'|x_{i+1})=P_F^f(b_i')$ and $P_B^i(b_i'|x_{i+1})=P_F^f(b_i'|x_i')$ respectively, where $P_B^i(b_i'|x_{i+1})$ denotes the initial distribution of the bit in the i-th period conditioned on the final state of the system in the (i+1)-th period in the backward process, $P_F^f(b_i')$ denotes the final distribution of the bit in the i-th period in the forward process, $P_F^f(b_i'|x_i')$ denotes the final distribution of the bit in the i-th period conditioned on the final state of the system in the i-th period in the forward process. Please note that the initial distribution of every bit is the same in the forward process, but it is not necessarily the same in the backward process. Corresponding to these two scenarios we can derive two different Fluctuation Theorems.
\subsubsection{CASE A: WITH THE DECORRELATION PROCESS AT THE END OF EVERY PERIOD}

We choose the initial distribution of bits in the backward process as the distribution of the outgoing bits in the forward process $P_B^i(b_i'|x_{i+1})\equiv P_F^f(b_i')$. The physical meaning of this choice of the backward process is the following, after finishing the forward process from the 0th to the nth bit, reverse the tape and let it go through the machine from the nth to the 0th bit, and what you get is just the backward process.

From the setting of the initial distribution of every bit in the backward process, one can obtain the ratio of probabilities between a pair of forward and backward trajectories,
\begin{equation}\label{equ10}
\begin{aligned}
	\frac{P_B\left(X_B,B_B\right)}{P_F\left(X_F,B_F\right)}=&\frac{P_F^f\left(x_n',b_n'\right)}{P_F^i\left(x_0,b_0\right)}\cdot \frac{P_B\left(X_{B,0},B_{B,0}|x_0',b_0'\right)}{P_F\left(X_{F,0},B_{F,0}|x_0,b_0\right)}\cdot\\&\frac{P_F^f\left(b_0'\right)}{P_F^i\left(b_1\right)}\cdot \frac{P_B\left(X_{B,1},B_{B,1}|x_1',b_1'\right)}{P_F\left(X_{F,1},B_{F,1}|x_1,b_1\right)}\cdot\\ &\frac{P_F^f\left(b_1'\right)}{P_F^i\left(b_2\right)}\cdot \cdot \cdot \cdot,
\end{aligned}
\end{equation}
where $P_B\left(X_B,B_B\right)$ and $P_F\left(X_F,B_F\right)$ represent the probability distributions of the backward and the forward trajectories of the composite system, $P_F^i\left(x_i,b_i\right)$ and $P_F^f\left(x_i',b_i'\right)$ represent the  probability distributions of the states of the composite system at the beginning and the end of the i-th period in the forward process, $P_F^i\left(b_i\right)$ and $P_F^f\left(b_i'\right)$ represent the marginal probability distributions of the states of the bit at the beginning and the end of the i-th period in the forward process, $P_F\left(X_{F,i},B_{F,i}|x_i,b_i\right)$ and $P_B\left(X_{B,i},B_{B,i}|x_i',b_i'\right)$ represent the probability distributions of the trajectories in the i-th period conditioned on the initial state of this period in the forward and the backward processes. From Eq.(\ref{equ10}), we can obtain
\begin{equation}\label{equ12}
\begin{aligned}
\left< e^{-\sum_{i=0}^{n}\left(\Delta s_{x,i}+\Delta \mathscr{I}_i\right)-\Delta s_{bath}}\right>=1,
\end{aligned}
\end{equation}
where $\Delta s_{x,i}$ and $\Delta \mathscr{I}_i$ represent the fluctuating entropy change of the system and the fluctuating information change of the bit in the i-th period. The detailed derivation of this equation can be found in Appendix A. This is one of the main results in our paper. It is an IFT for a class of information machines. We would like to emphasize that this IFT is very general. It is valid for an arbitrary initial distribution of the system and an arbitrary initial distribution of the bit flow. Also, it is valid for an arbitrary number of periods. Using Jensen's Inequality we can immediately obtain the second law,
\begin{equation}\label{equ14}
	\sum_{i=0}^{n}\left(\Delta S_{x,i}+\Delta I_{b,i}\right)+\Delta S_{bath}\geq 0,
\end{equation}
where $\Delta S_{x,i}$ and $\Delta I_{b,i}$ represent the ensemble averaged entropy change of the system and the information change of the bit in the i-th period.

Please notice that there is a difference between the LHS of Eq.(\ref{equ14}) and the total entropy production (see Eq.(\ref{equ20})), which is the sum of the mutual entropies between the system and the bit at the end of each period. One can understand the physical process underlying Eq.(\ref{equ14}) as follows: A decorrelation process between the system and the bit occurs before it interacts with a new bit. That is, there is an entropy increment following the dynamic evolution of the composite system, which is denoted by the dashed arrow in Fig. \ref{fig4}. This imaginary decorrelation process is reasonable. We remember that before the system interacts with the first bit, the system and the tape are uncorrelated, so before it interacts with the latter bits, it could be uncorrelated with the tape either. We would like to emphasize that, the imaginary decorrelation processes do not influence the work extraction in the information engine, but they influence the total entropy production.

What's more, if we consider the periodic steady states, the term $\Delta S_{x,i}$ vanishes, and we obtain
\begin{equation}\label{equ16}
	T\cdot\sum_{i=0}^{n}\left(\Delta I_{b,i}\right)+Q\geq 0,
\end{equation}
where $\frac{Q}{T}=\Delta S_{bath}$ is used. This is Laudauer's principle~\cite{Landauer1961IBM}.

We can also consider a DFT. If the initial distribution of the backward process is equal to the final distribution of the forward process, by summing up the probabilities of those trajectories with the same total entropy production, we can rewrite Eq.(\ref{equ10}) as
\begin{equation}\label{equ17}
	\frac{P_F\left(\Delta s_{tot}\right)}{P_B\left(-\Delta s_{tot} \right)}=e^{\Delta s_{tot}},
\end{equation}
where $\Delta s_{tot}=\sum_{i=0}^{n}\left(\Delta s_{x,i}+\Delta \mathscr{I}_i\right)+\Delta s_{bath}$, and $P_F(\Delta s_{tot})$ and $P_B(\Delta s_{tot})$ represent the probability distributions of the total entropy production in the forward and the backward processes respectively. Eq.(\ref{equ17}) is the DFT in our models.

We would like to emphasize that the DFT in steady states proposed by Seifert (see Eq.(21) of Ref. \cite{Seifert2005PRL}) does not exist in the periodic steady states of our current information machine. The reason is that in the periodic steady states, the distribution of the system does not change in each period, but the distribution of the bit changes.
\subsubsection{CASE B: WITHOUT THE DECORRELATION PROCESS AT THE END OF EVERY PERIOD}
In Case A,  the total entropy change in the Fluctuation Theorems (Eq.(\ref{equ12}) and Eq.(\ref{equ17})) does not contain the mutual entropy at the end of every period between the system and every bit. In this section, we would like to derive the IFT and the DFT related to the real total entropy production.

If in the forward process there is no decorrelation process at the end of every period, in the backward process, the probability distribution of the incoming bit is given by $P_B^i(b_i'|x_{i+1})\equiv P_F^f(b_i'|x_i')$. Notice that now the transition probability of the backward process $P_B^i(b_i'|x_{i+1})$ not only depends on the initial condition, as is the case in Case A, but also depends on the final state of the system in the i-th period in the forward process $x_i'$. Because we have set $x_i'=x_{i+1}$, the backward process can be constructed in the following way, at the end of every period, the probability distribution of the next bit will depend on the state of the system $x_{i+1}$ which is $P_B^i(b_i'|x_{i+1})\equiv P_F^f(b_i'|x_{i+1})=P_F^i(b_i'|x_i')$. This choice implies that for different state sequence of the system $\{x_i\}$, the choices of the initial distribution of the bits in the backward process are different. However, in Case A, the choice of the initial distribution of the bits does not depend on the state sequence. For this reason, we should subgroup the trajectories in the backward process according to the the state sequence $\{x_i\}$. The physical meaning of this choice is as follows, suppose that we have an ensemble of tapes already gone through the machine, then we pick up the ones that have the same $\{x_i\}$ sequence to form a new ensemble, and let them go through the machine reversely. For different subgroups $\{x_i\}$, $\{P_B^i(b_i'|x_{i+1})\}$ are different.

Now, we can similarly derive an IFT as follows,
\begin{equation}\label{equ18}
\begin{aligned}
	\frac{P_B\left(X_B,B_B\right)}{P_F\left(X_F,B_F\right)}=&\frac{P_F^f\left(x_n',b_n'\right)}{P_F^i\left(x_0)\cdot P_F^i(b_0\right)}\cdot \frac{P_B\left(X_{B,0},B_{B,0}|x_0',b_0'\right)}{P_F\left(X_{F,0},B_{F,0}|x_0,b_0\right)}\cdot\\ &\frac{P_F^f\left(b_0'|x_0'\right)}{P_F^i\left(b_1\right)}\cdot \frac{P_B\left(X_{B,1},B_{B,1}|x_1',b_1'\right)}{P_F\left(X_{F,1},B_{F,1}|x_1,b_1\right)}\cdot\\& \frac{P_F^f\left(b_1'|x_1'\right)}{P_F^i\left(b_2\right)}\cdot \cdot \cdot \cdot\\
	=&e^{-\Delta s_{bath}}\cdot\frac{P_F^f(x_n')}{P_F^i(x_0)}\cdot \prod_{i=0}^{n}\frac{P_F^f(b_i'|x_i')}{P_F^i(b_i)}\\
	=&e^{-\Delta s_{bath}}\cdot \prod_{i=0}^{n}\frac{P_F^f(b_i',x_i')}{P_F^i(b_i)\cdot P_F^i(x_i)}\\
	=&e^{-\Delta s_{bath}-\sum_{i=0}^{n}(\Delta s_{x,i}+\Delta \mathscr{I}_i-\Delta \mathscr{M}_i)},
\end{aligned}
\end{equation}
where $P_F^i(x_i)$ and $P_F^f(x_i')$ represent the marginal probability distributions of the states of the system at the beginning and the end of the i-th period in the forward process, $\Delta \mathscr{M}_i$ represents the fluctuating mutual entropy change between the system and the bit in the i-th period. And the detailed balance condition Eq.(\ref{a2}) is used. After taking the average, we obtain the IFT,
\begin{equation}\label{equ19}
<e^{-\Delta s_{bath}-\sum_{i=0}^{n}(\Delta s_{x,i}+\Delta \mathscr{I}_i-\Delta \mathscr{M}_i)}>=1,
\end{equation}
from which we can also derive the second law,
\begin{equation}\label{equ20}
	\Delta S_{bath}+\sum_{i=0}^{n}(\Delta S_{x,i}+\Delta I_i-\Delta M_i)\geq 0,
\end{equation}
where $\Delta M_i$ represents the mutual entropy change between the system and the bit in the i-th period. This relation also appears in the supplemental material of Ref. \cite{Mandal2013PRL}.
Notice that $\Delta M_i\geq0$, so Eq.(\ref{equ20}) is tighter than Eq.(\ref{equ14}). Similar to Case A, we also have the DFT,
\begin{equation}\label{equ21}
\frac{P_F\left(\Delta s_{tot}\right)}{P_B\left(-\Delta s_{tot} \right)}=e^{\Delta s_{tot}},
\end{equation}
where $\Delta s_{tot}=\sum_{i=0}^{n}(\Delta s_{x,i}+\Delta\mathscr{I}_i-\Delta \mathscr{M}_i)+\Delta s_{bath}$. Similar to Case A, the relations Eq.(\ref{equ19}), Eq.(\ref{equ20}), Eq.(\ref{equ21}) are valid for arbitrary initial states of the system and the bit. If the system has reached the periodic steady states, one can find a tighter relation than the Landauer's principle Eq.(\ref{equ16}) for this model.
\subsection{NUMERICAL RESULTS OF THE TOTAL ENTROPY PRODUCTION}
\begin{figure*}[p]
	\centering
	\subfigure[]{
		\includegraphics[width=0.48\textwidth]{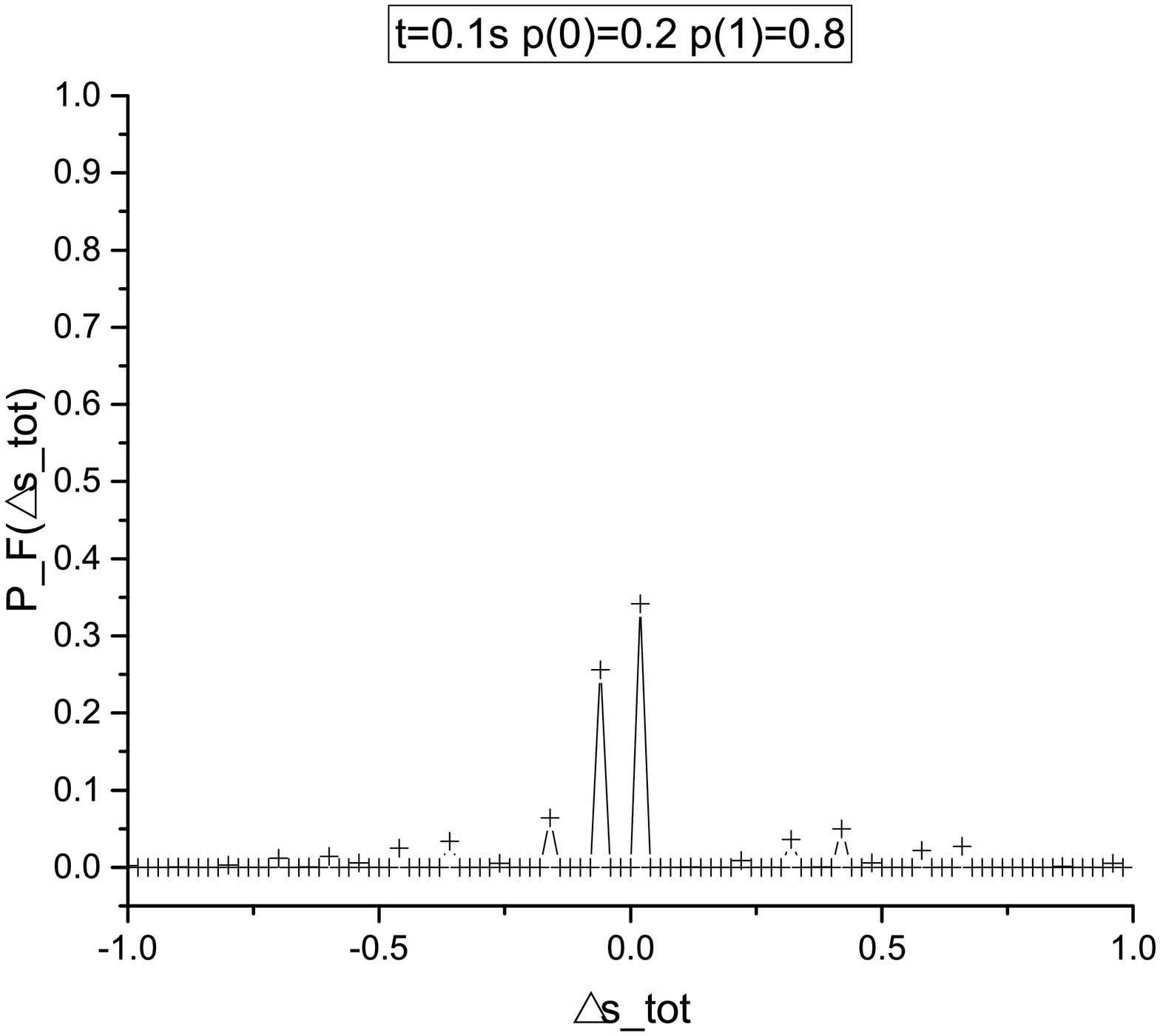}
		\label{fig5:a}}
	\subfigure[]{
			\includegraphics[width=0.48\textwidth]{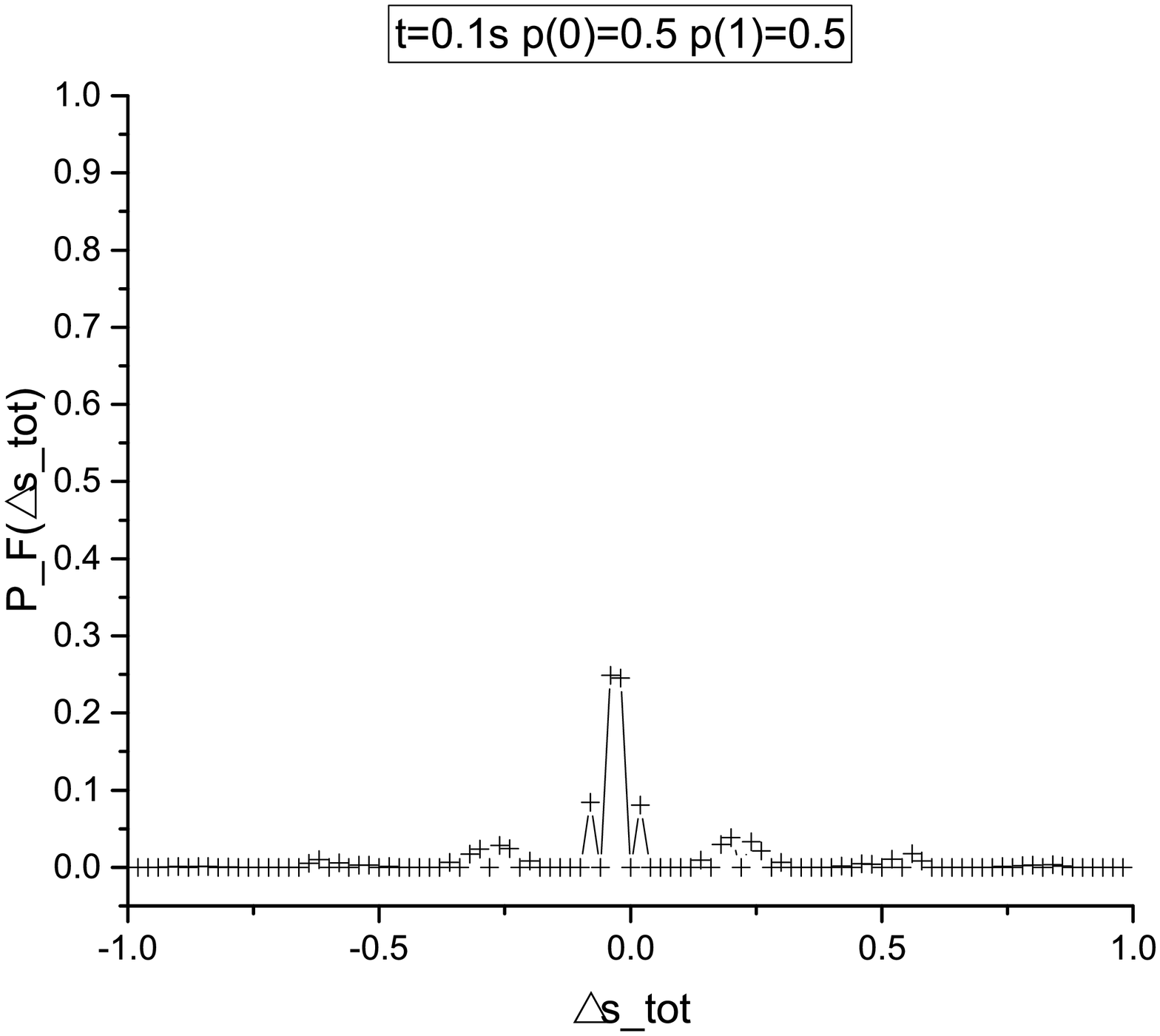}
			\label{fig5:b}}
	\subfigure[]{
		\includegraphics[width=0.48\textwidth]{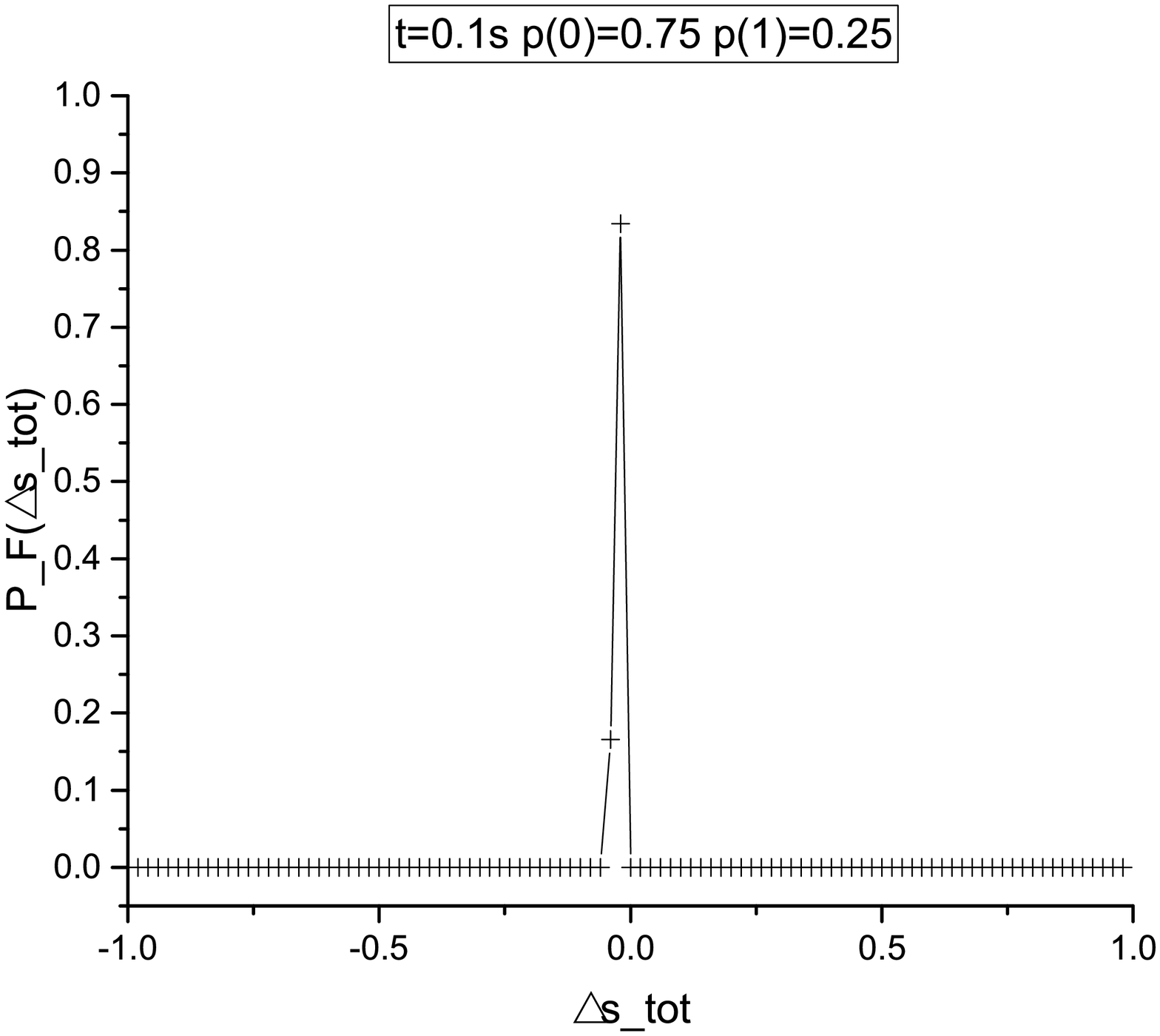}
		\label{fig5:c}}
	\subfigure[]{
			\includegraphics[width=0.48\textwidth]{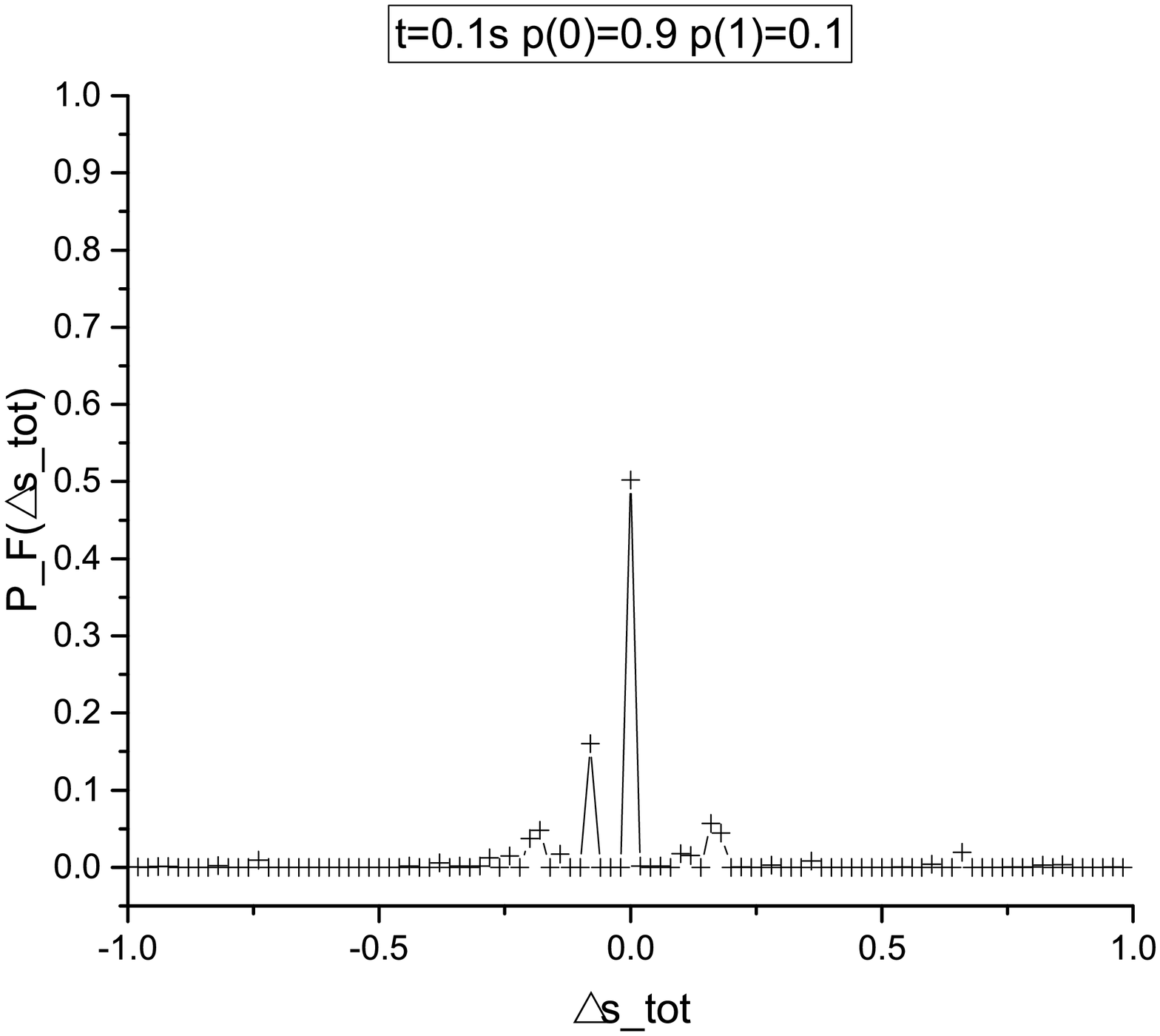}
			\label{fig5:d}}
	\subfigure[]{
			\includegraphics[width=0.48\textwidth]{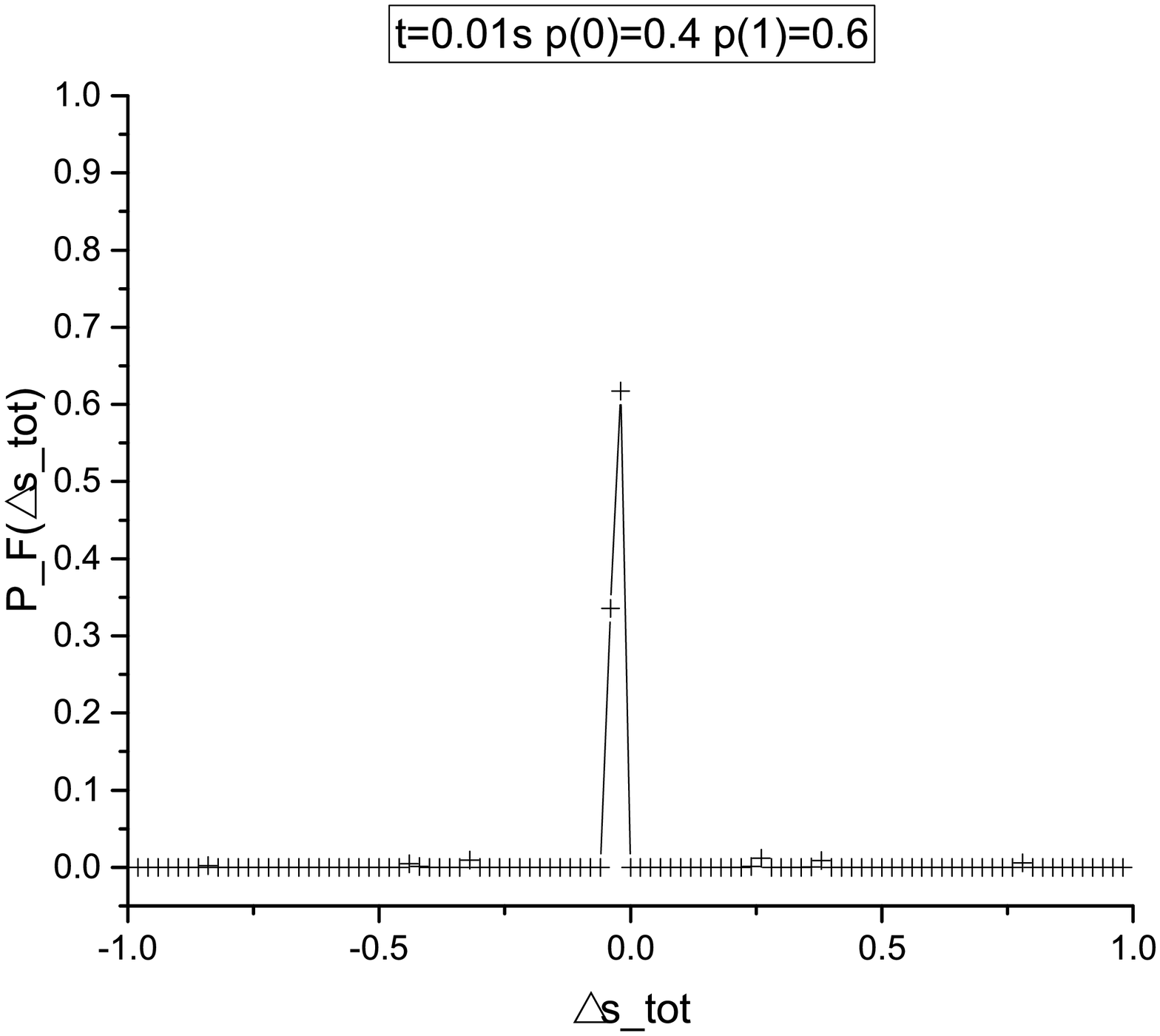}
			\label{fig5:e}}
	\subfigure[]{
			\includegraphics[width=0.48\textwidth]{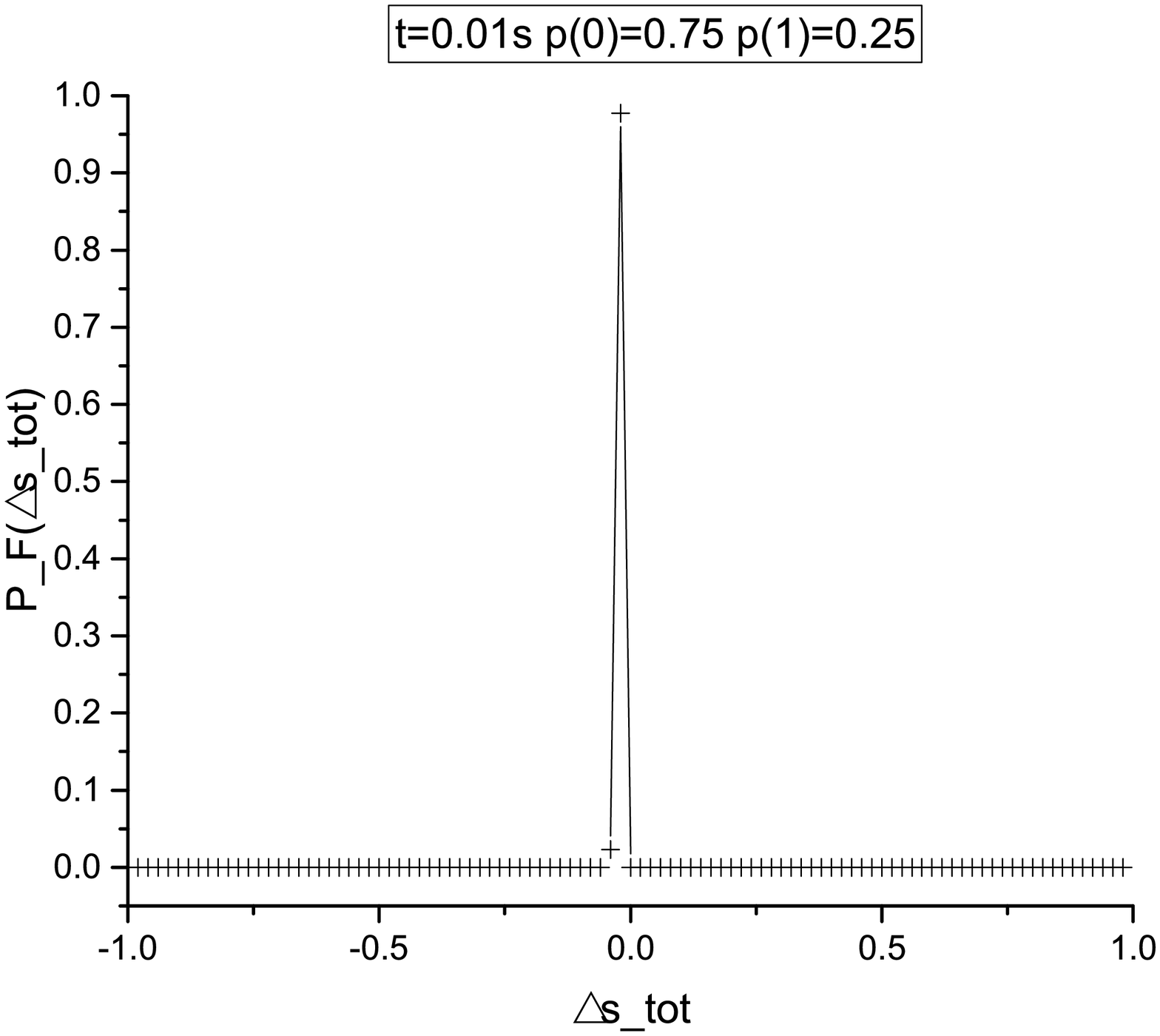}
			\label{fig5:f}}
	\caption{Probability distributions of the total entropy production in the periodic steady states in CASE A. Here, $\Delta s_{tot}=\sum_{i=0}^{n}\left(\Delta s_{x,i}+\Delta \mathscr{I}_i\right)+\Delta s_{bath}$. We have chosen the parameters to be $\epsilon=0.5$, n=2, and $(t, p(0), p(1))=(0.1s, 0.2, 0.8), (0.1s, 0.5, 0.5), (0.1s, 0.75, 0.25), (0.1s, 0.9, 0.1), (0.01s, 0.4, 0.6), (0.01s, 0.75, 0.25)$.}
	\label{fig5}
\end{figure*}

\begin{figure*}[p]
	\centering	
		\subfigure[]{
		\includegraphics[width=0.48\textwidth]{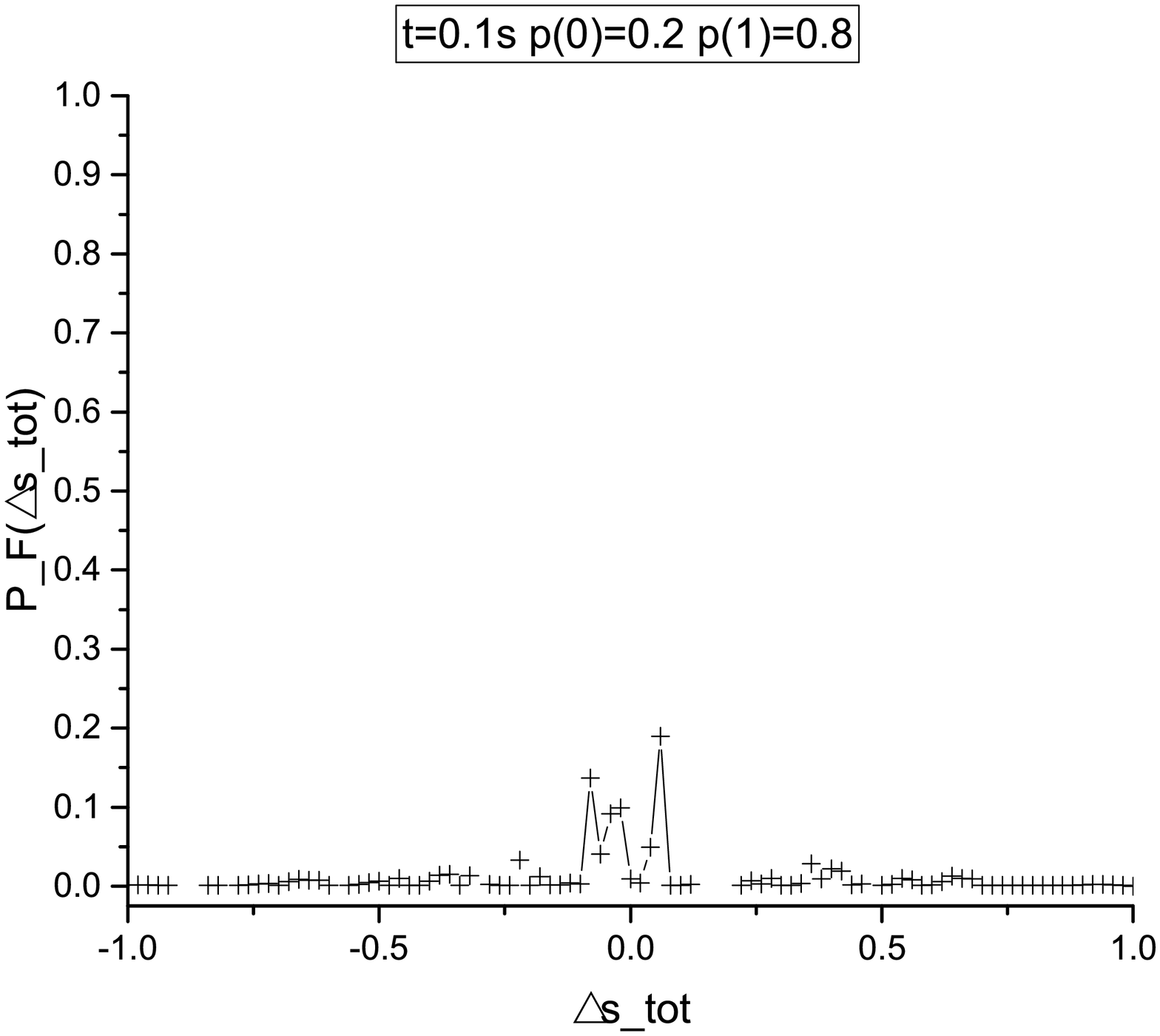}
		\label{fig6:a}}
	\subfigure[]{
		\includegraphics[width=0.48\textwidth]{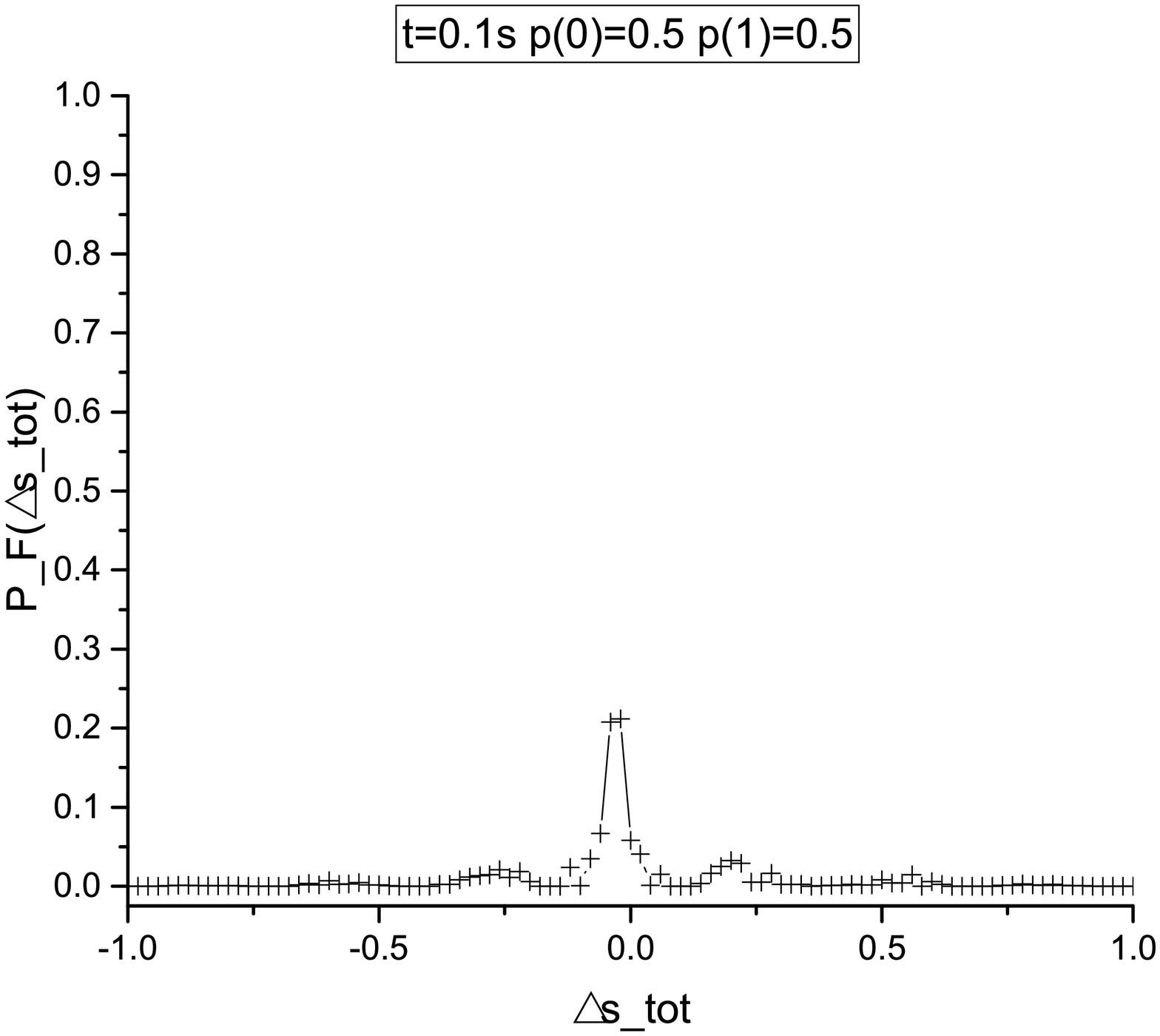}
		\label{fig6:b}}
		\subfigure[]{
		\includegraphics[width=0.48\textwidth]{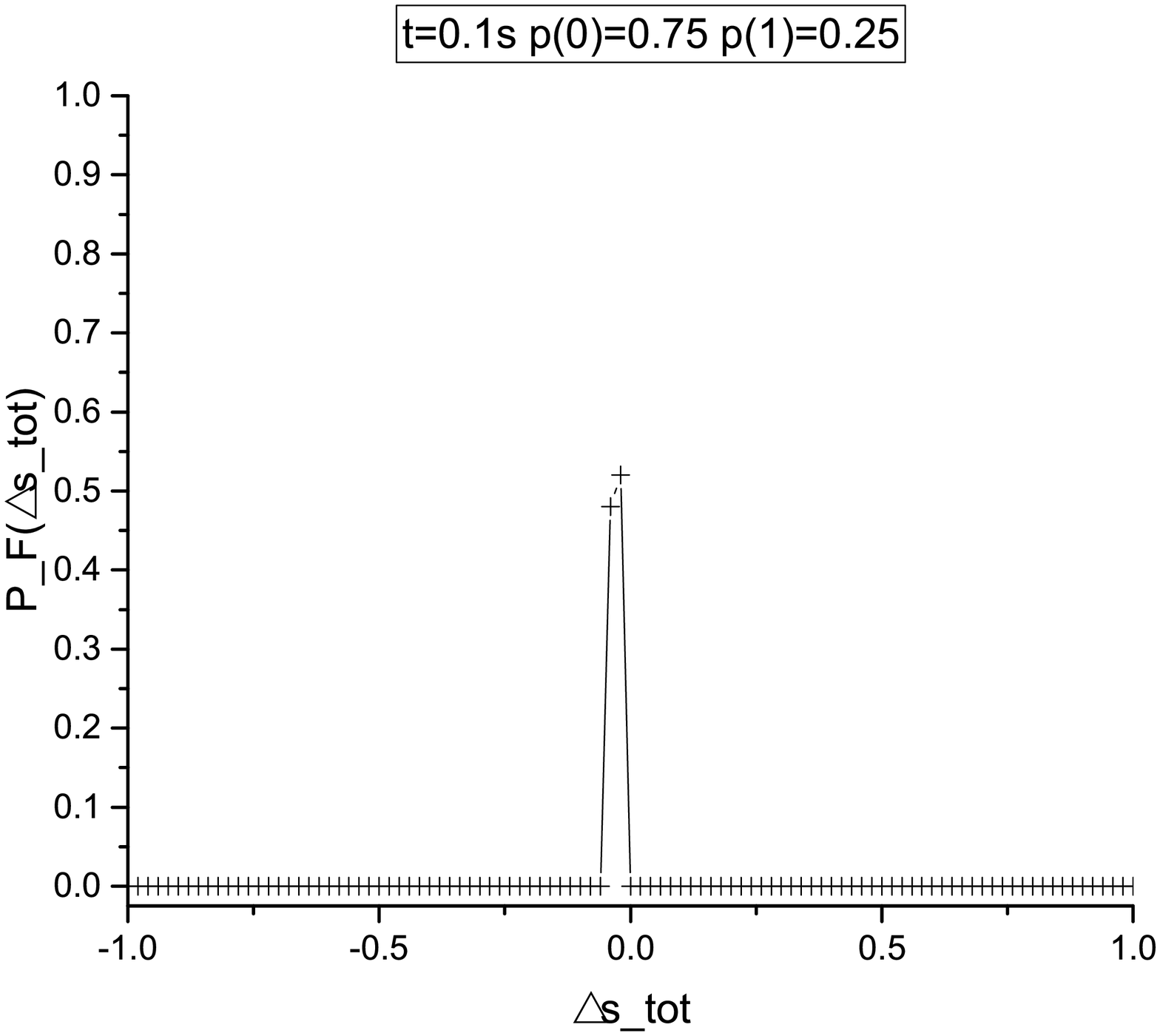}
		\label{fig6:c}}
	\subfigure[]{
		\includegraphics[width=0.48\textwidth]{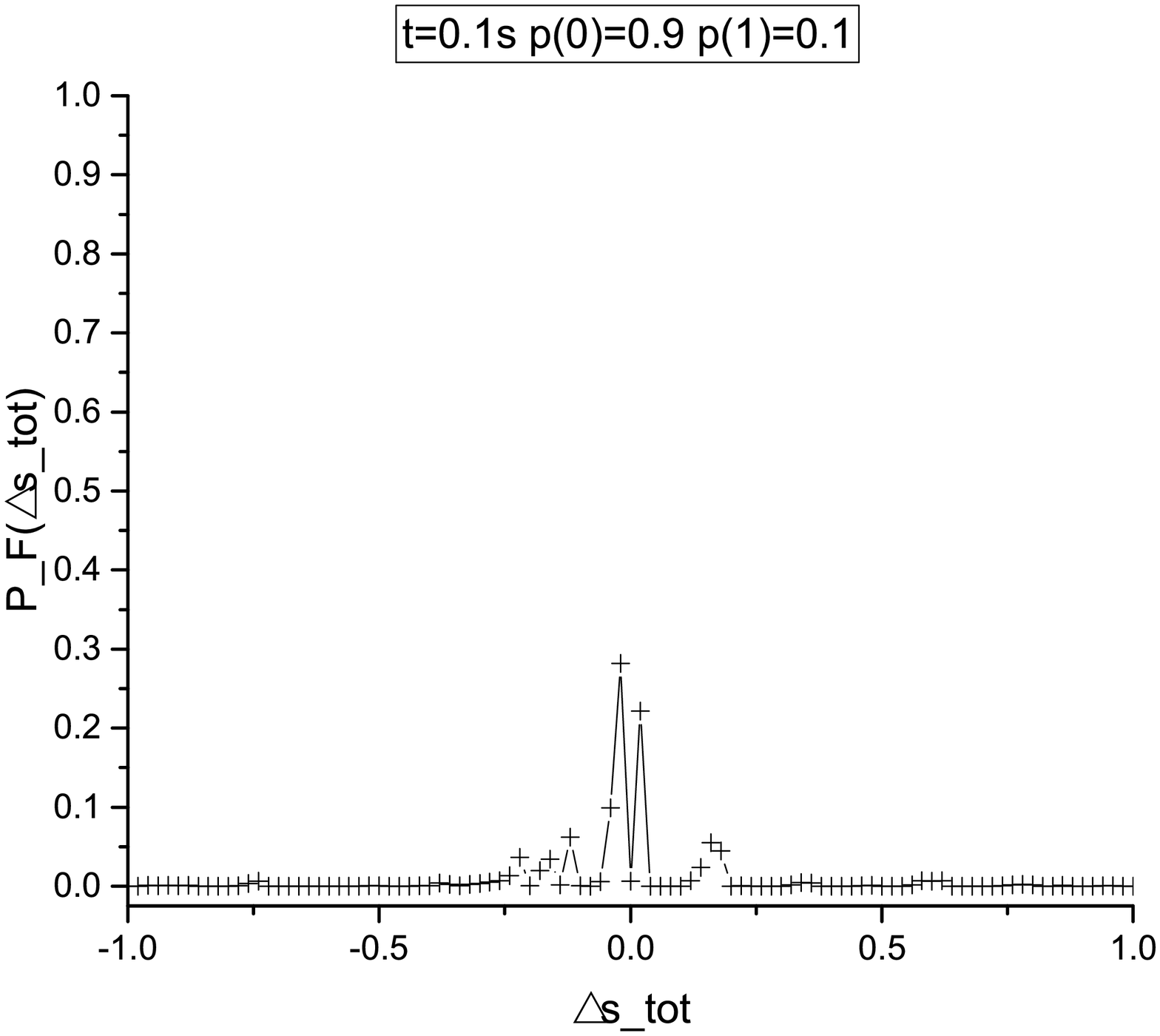}
		\label{fig6:d}}
		\subfigure[]{
		\includegraphics[width=0.48\textwidth]{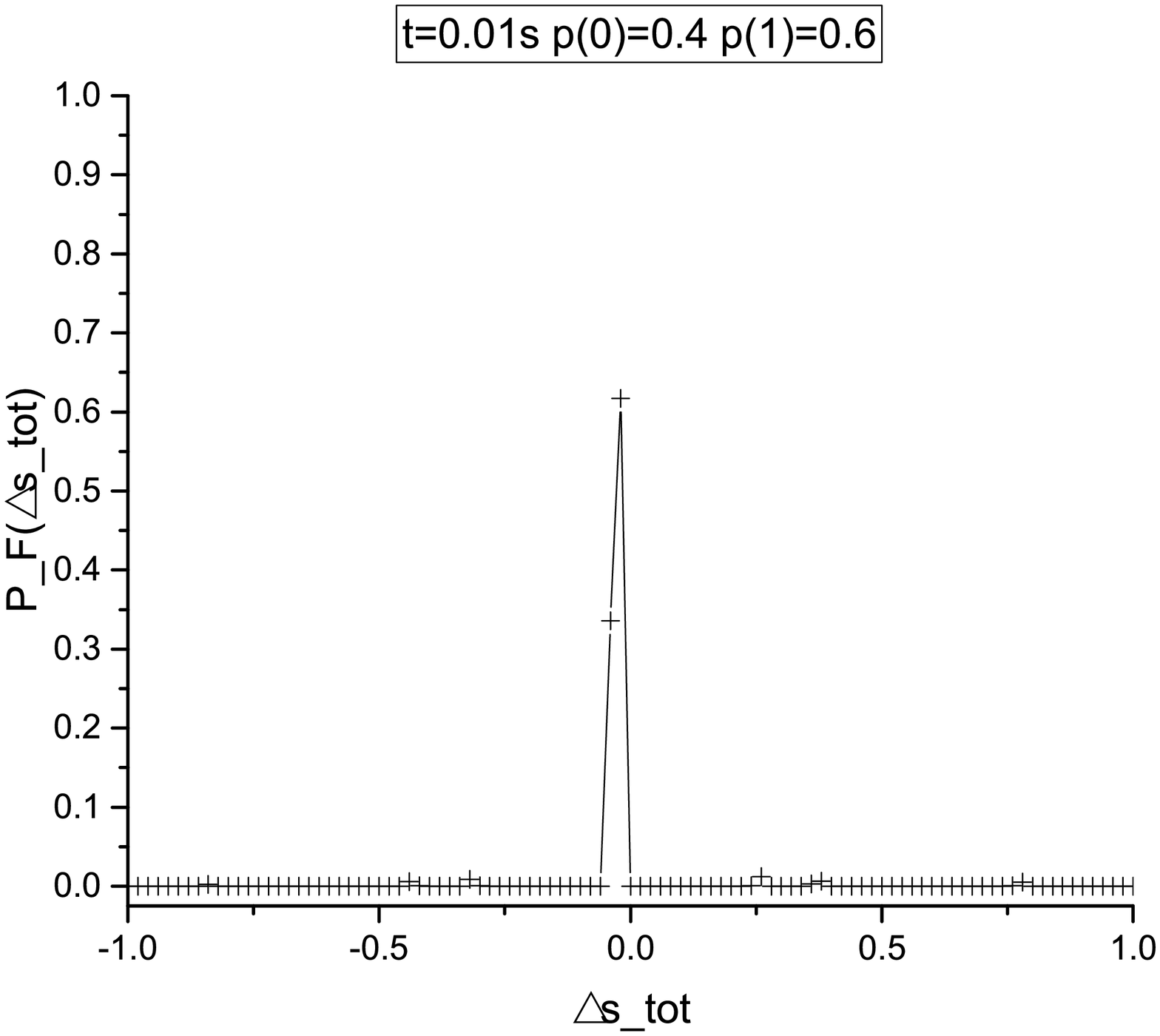}
		\label{fig6:e}}
	\subfigure[]{
		\includegraphics[width=0.48\textwidth]{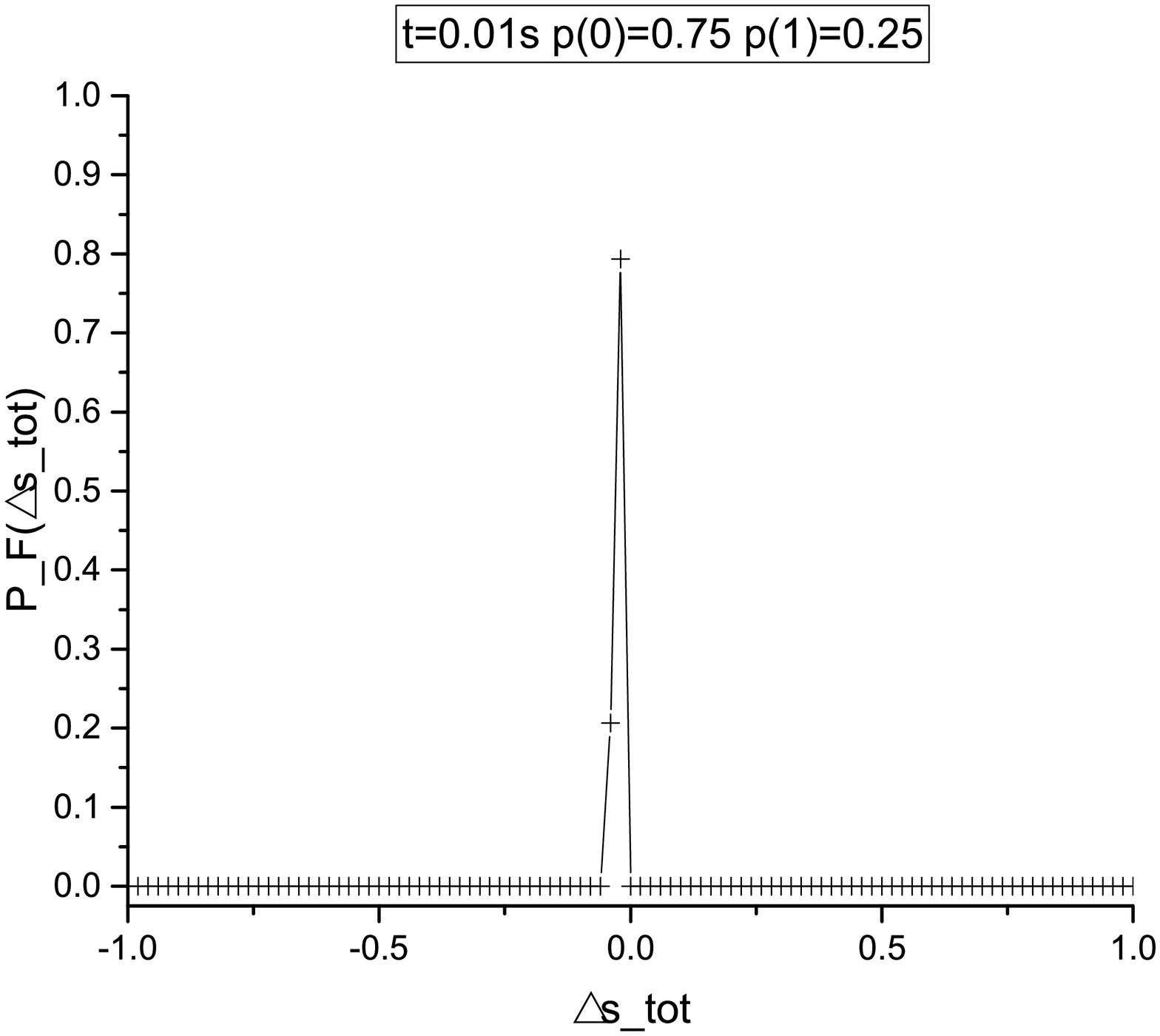}
		\label{fig6:f}}
	\caption{Probability distributions of the total entropy production in the periodic steady states in CASE B. Here, $\Delta s_{tot}=\sum_{i=0}^{n}\left(\Delta s_{x,i}+\Delta \mathscr{I}_i-\Delta\mathscr{M}_i\right)+\Delta s_{bath}$. We have chosen the parameters to be $\epsilon=0.5$, n=2, and $(t, p(0), p(1))=(0.1s, 0.2, 0.8), (0.1s, 0.5, 0.5), (0.1s, 0.75, 0.25), (0.1s, 0.9, 0.1), (0.01s, 0.4, 0.6), (0.01s, 0.75, 0.25)$.}
	\label{fig6}
\end{figure*}

In order to show the results above visually, we calculate the probability distributions of the total entropy production in the two IFTs (Eq.(\ref{equ12}), Eq.(\ref{equ19})) numerically. The model we choose is the information engine~\cite{Mandal2012PNAS}, whose transition rate matrix $W_{ij}$ is given by
\begin{equation}\label{equ22}
\left(
	\begin{matrix}
	-1 &1 &0 &0 &0 &0\\
	1 &-2 &1 &0 &0 &0\\
	0 &1 &-2+\epsilon &1+\epsilon &0 &0\\
	0&0&1-\epsilon &-2-\epsilon &1&0\\
	0&0&0&1&-2&1\\
	0&0&0&0&1&-1\\
	\end{matrix}
\right),
\end{equation}
where $\epsilon\equiv p(0)^{eq}-p(1)^{eq}=\tanh\left(\frac{mg\Delta h}{2T} \right)$, where $p(b)^{eq}$ denotes the marginal distribution of the bit in equilibrium, $mg\Delta h$ denotes the energy difference between the $b=0$ states and the $b=1$ states, T denotes the temperature of the heat bath. In our calculation, we choose $\epsilon=0.5$ and we consider a process that consists of three periods in the periodic steady states. The results are illustrated in Fig. \ref{fig5} and Fig. \ref{fig6}.

From these figures, we can see that when $p(0)=p(0)^{eq}=0.75$ and $p(1)=p(1)^{eq}$=0.25, the distributions can reach a delta distribution centered at $\Delta s_{tot}=0$ (see Fig. \ref{fig5:c}, Fig. \ref{fig5:f}, Fig. \ref{fig6:c} and Fig. \ref{fig6:f}).

This result can be understood as follows. If the initial distribution of the states of the bits is equal to the equilibrium distribution, the distribution of the six states will not change under the evolution governed by the Master equation. So that the system and the bit will not be correlated $\Delta\mathscr{M}_i=0$. What's more, the probability distribution of the three states will be just equal to each other. $\Delta s_{bath}$ cancels out with $\sum_{i=0}^{n}\Delta \mathscr{I}_i$. So the changes of the fluctuating total entropy will always be equal to 0 in both cases.

The length of the interacting interval also influences the distributions of the fluctuating total entropy production in the periodic steady states. The shorter the interacting interval of every period t is, the closer are the distributions of the fluctuating entropy production to the delta distribution. This result can be understood as follows, the shorter is the time interval of every period, the less will the heat bath influence the system and the bit.
\section{APPLICATION TO A NEW MODEL}
In this section, we will exploit our previous methods to study a new information-device model proposed by Thomas McGrath, et al. recently~\cite{Mcgrath2016PRL}. Their model is about chemical reactions, polymers and enzymes. But the mathematical essence of their device is quite similar to the model which we discuss in Sec. II. The main difference is that it has two tapes instead of one.
\begin{figure}[!htb]
	\centering	
		\subfigure[]{
		\includegraphics[width=0.2\textwidth]{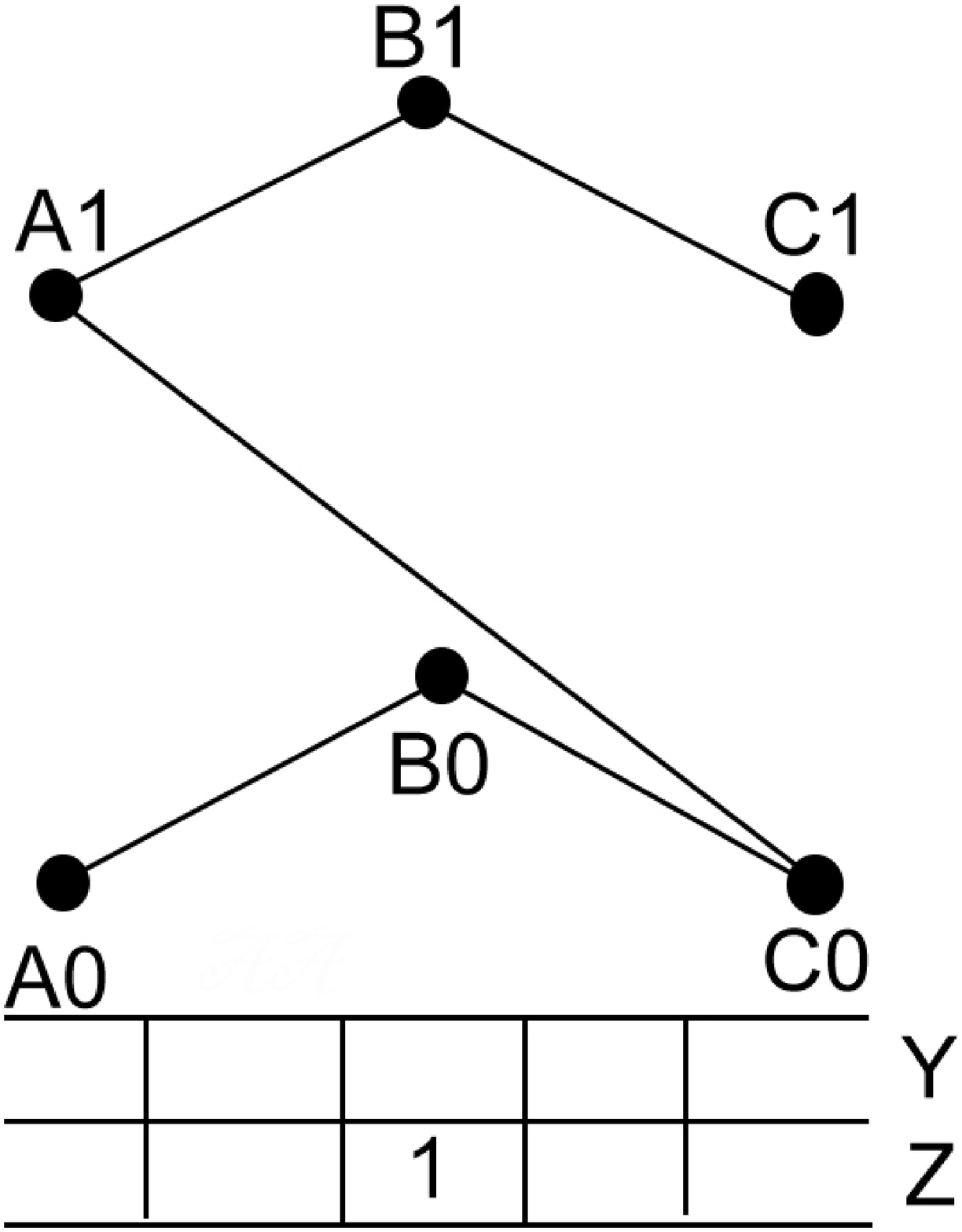}
		\label{}}
	    \subfigure[]{
		\includegraphics[width=0.2\textwidth]{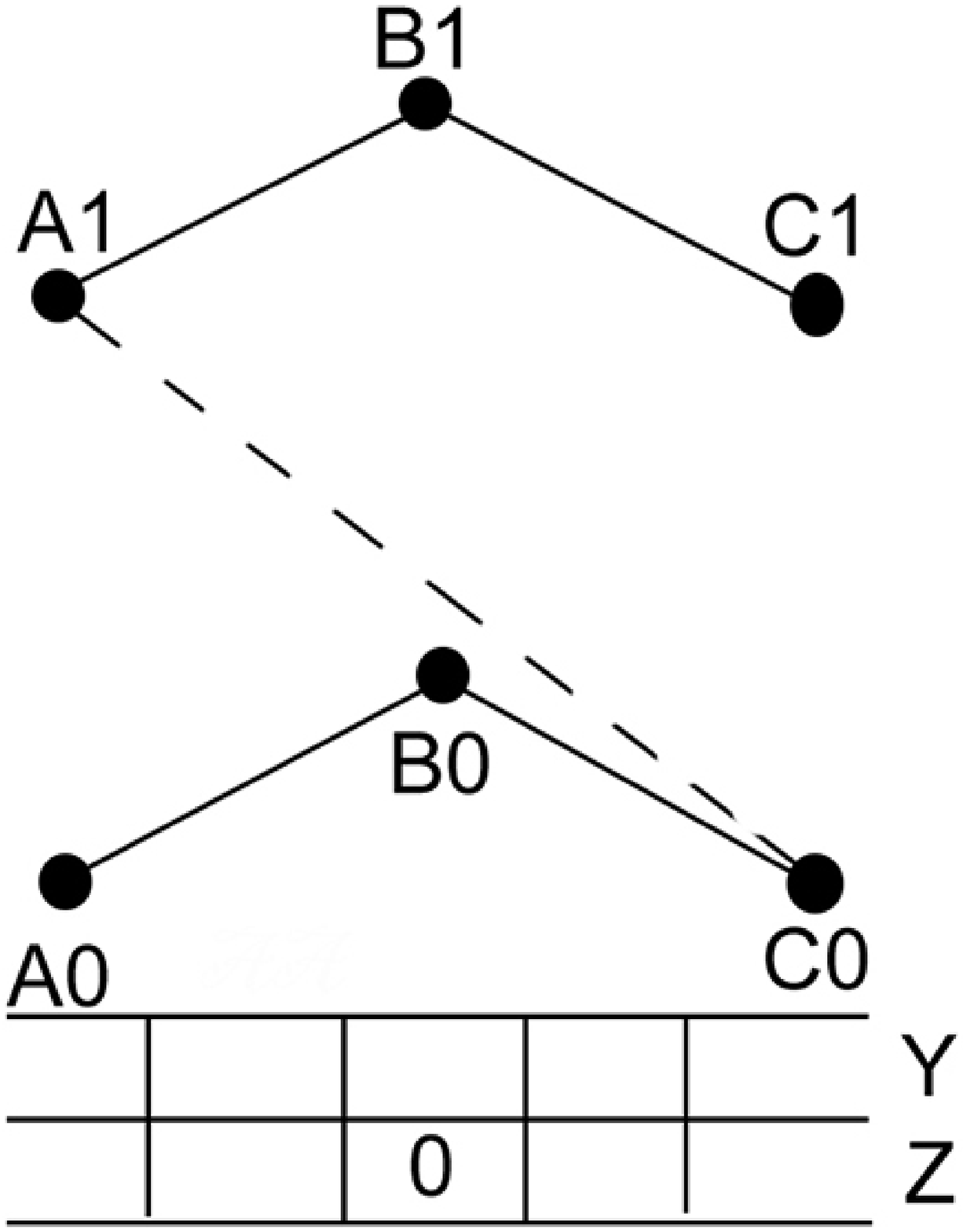}
		\label{}}
\caption{Schematic plots of the model proposed in Ref.~\cite{Mcgrath2016PRL}. The solid lines denote the transitions that are allowed, the dashed line denotes the transition that is forbidden. A, B and C represent the 3 states of the system. 0 and 1 denote the two states of the bit. Y and Z denote the two tapes.}
\label{fig7}
\end{figure}

As shown in Fig. \ref{fig7}, tape Y is just the same as the tape in the previous models, while tape Z is an extra one. In the process, the bit of tape Z does not change its value, and its function is to control the transition paths. When the value of the bit of tape Z passing through the system is 1, the allowed transitions will be just the same as before shown in Fig. \ref{fig2:a}. On the contrary, when the value is 0, the transition between C0 and A1 will be forbidden.

This model is interesting when there is correlation between tape Y and tape Z. For example, when these two tapes are fully correlated and the values of the bits are complementary. That is, if the bit in tape Y is 1 (0), the accompanied bit in tape Z must be 0 (1). In this situation, one can see that  in every single period the device can only do positive work or zero work. It can never do negative work, because if the initial value of tape Y is 1, the transition between C0 and A1 will be forbidden so that the system cannot go down to the b=0 states to do negative work. Therefore, the total work increases at the price of (partially) destroying the correlation between the two tapes.

Now, we can study the Fluctuation Theorem for this model. First, we can view these two tapes as one tape. This tape consists of many pairs of bits. Every pair of bits has 4 possible values: 00, 01, 10 and 11. With this in mind, the analysis and the results of the Fluctuation Theorem for this model are the same as those in Sec. II. So Eq.(\ref{equ19}), Eq.(\ref{equ20}) and Eq.(\ref{equ21}) also hold true for this model.

Furthermore, we can decompose the total entropy production of the tape in this way. The fluctuating and the ensemble averaged information entropy of every pair of bits can be expressed as $\Delta\mathscr{I}_i=\Delta\mathscr{I}_i^{Y}+\Delta\mathscr{I}_i^{Z}-\Delta\mathscr{M}_i^{Y,Z}$ and $\Delta I_{b,i}=\Delta I_{b,i}^{Y}+\Delta I_{b,i}^{Z}-\Delta M_i^{Y,Z}$, where $\Delta\mathscr{I}_i^{Y}$, $\Delta\mathscr{I}_i^{Z}$ and $\Delta\mathscr{M}_i^{Y,Z}$ denote the fluctuating entropy changes of tape Y, tape Z, and the fluctuating mutual entropy change between them,  $\Delta I_{b,i}^{Y}=<\Delta\mathscr{I}_i^{Y}>$, $\Delta I_{b,i}^{Z}=<\Delta\mathscr{I}_i^{Z}>$ and $\Delta M_i^{Y,Z}=<\Delta\mathscr{M}_i^{Y,Z}>$  denote the entropy productions of tape Y, the tape Z, and the mutual entropy change between them. Because the values of the bits in tape Z do not change, $\Delta \mathscr{I}_i^{Z}=\Delta I_{b,i}^{Z}=0$. Then Eq.(\ref{equ19}) and Eq.(\ref{equ20}) become
\begin{equation}\label{equ27}
\left< e^{-\sum_{i=0}^{n}\left(\Delta s_{x,i}+\Delta \mathscr{I}_i^Y-\Delta \mathscr{M}_i^{Y,Z}-\Delta \mathscr{M}_i^{X,YZ}\right)-\Delta s_{bath}}\right>=1,
\end{equation}
\begin{equation}\label{equ28}
\sum_{i=0}^{n}\left(\Delta S_{x,i}+\Delta I_{b,i}^Y-\Delta M_{i}^{Y,Z}-\Delta M_i^{X,YZ}\right)+\Delta S_{bath}\geq 0,
\end{equation}
where $\Delta \mathscr{M}_i^{X,YZ}$ denotes the fluctuating mutual entropy change between the system and the composite tape Y and Z, and $\Delta M_{i}^{X,YZ}$ denotes the mutual entropy change between the system and the composite tape Y and Z.

Now, we can see that the device can do work by exploiting not only the information entropy, but also the mutual entropy between the two tapes. We would like to emphasize that in many previous studies of Fluctuation Theorems containing information, the information content is about the mutual entropy only~\cite{Sagawa2010PRL, Sagawa2012PRL}. But in our current study, the information content includes both the information entropy and the mutual entropy.

Actually, we can view tape Z as a measurement-feedback device, measuring the initial state of tape Y in every period, and then doing a feedback to control the transition between C0 and A1. If the tapes are fully correlated (so the mutual entropy is just equal to the entropy of one tape), the corresponding measurement will be viewed as a perfect measurement. Otherwise the measurement will be viewed as imperfect. Therefore, we can view this device as a combination of a low-entropy-tape-fueled device~\cite{Mandal2012PNAS} and a measurement-feedback device~\cite{Sagawa2010PRL,Sagawa2012PRL}.
\section{CONCLUSION}
In summary, we have derived the FTs containing information and the second law for a class of information machines which can be viewed as autonomous Maxwell's demon-assisted machines. Because the mutiple-bits process in our information machine cannot be described by a single Master equation, the FTs can not be regarded as  special cases of the standard FT~\cite{Seifert2005PRL}. From our IFT, we have straightforwardly obtained the Landauer's principle Eq.(\ref{equ16}). Finally we studied an application of our results to a new information device.

In the future it might be interesting to explore whether the results also hold true for the quantum version. Because in quantum mechanics, the entropy of entanglement is very similar to the mutual entropy in classical cases.

\begin{acknowledgements}
H. T. Quan gratefully acknowledges support from the National Science Foundation of China under grants 11375012, 11534002, and The Recruitment Program of Global Youth Experts of China.
\end{acknowledgements}
\appendix
\section{Derivation of Eq.(\ref{equ12})}
Because $P_F^f\left(x_n',b_n'\right)=P_F^f\left(x_n'\right)\cdot P_F^f\left(b_n'\right)\cdot \frac{P_F^f\left(x_n',b_n'\right)}{P_F^f\left(x_n'\right)\cdot P_F^f\left(b_n'\right)}$ and $P_F^i\left(x_0,b_0\right)=P_F^i\left(x_0\right)\cdot P_F^i\left(b_0\right)$, Eq.(\ref{equ10}) can be written as
\begin{equation}\label{equ11}
\begin{aligned}
&P_B\left(X_B,B_B\right)\cdot\frac{P_F^f\left(x_n'\right)\cdot P_F^f\left(b_n'\right)}{P_F^f\left(x_n',b_n'\right)}\\=&P_F\left(X_F,B_F\right)\cdot \prod_{i=0}^{n}\frac{P_F^f\left(b_i'\right)}{P_F^i\left(b_i\right)}\cdot\prod_{i=0}^{n}\frac{P_F^f\left(x_i'\right)}{P_F^i\left(x_i\right)}\cdot\\& \prod_{i=0}^{n}\frac{P_B\left(X_{B,i},B_{B,i}|x_i',b_i'\right)}{P_F\left(X_{F,i},B_{F,i}|x_i,b_i\right)}\\=&P_F\left(X_F,B_F\right)\cdot e^{-\sum_{i=0}^{n}\left(\Delta s_{x,i}+\Delta \mathscr{I}_i\right)-\Delta s_{bath}},
\end{aligned}
\end{equation}
where we have used the detailed balance condition
 \begin{equation}\label{a2}
 \frac{P_B\left(X_{B,i},B_{B,i}|x_i',b_i'\right)}{P_F\left(X_{F,i},B_{F,i}|x_i,b_i\right)}=e^{-\Delta s_{bath,i}},
 \end{equation}
 where $\Delta s_{bath,i}$ denotes the fluctuating entropy change of the heat bath in the i-th period. Then, we can calculate the summation of both sides of Eq.(\ref{equ11}),
 ~\\
 ~\\
 ~\\
\begin{equation}
\begin{aligned}
&\left< e^{-\sum_{i=0}^{n}\left(\Delta s_{xi}+\Delta \mathscr{I}_i\right)-\Delta s_{bath}}\right>\\=&\sum_{X_F, B_F} P_F\left(X_F,B_F\right)\cdot e^{-\sum_{i=0}^{n}\left(\Delta s_{xi}+\Delta \mathscr{I}_i\right)-\Delta s_{bath}}\\
=&\sum_{X_F, B_F} 	P_B\left(X_B,B_B\right)\cdot\frac{P_F^f\left(x_n'\right)\cdot P_F^f\left(b_n'\right)}{P_F^f\left(x_n',b_n'\right)}\\
=&\sum_{X_B, B_B} 	P_B\left(X_B,B_B\right)\cdot\frac{P_F^f\left(x_n'\right)\cdot P_F^f\left(b_n'\right)}{P_F^f\left(x_n',b_n'\right)}\\
=&\sum_{X_B, B_B} 	P_B\left(X_B,B_B|x_n',b_n'\right)\cdot P_F^f\left(x_n',b_n'\right)\cdot\frac{P_F^f\left(x_n'\right)\cdot P_F^f\left(b_n'\right)}{P_F^f\left(x_n',b_n'\right)}\\
=&\sum_{X_B, B_B} 	P_B\left(X_B,B_B|x_n',b_n'\right)\cdot P_F^f\left(x_n'\right)\cdot P_F^f\left(b_n'\right)\\
=&1,
\end{aligned}
\end{equation}
which is Eq.(\ref{equ12}).
~\\
~\\
~\\
~\\
~\\

\end{document}